\documentclass{article}

\usepackage{microtype}
\usepackage{graphicx}
\usepackage{booktabs} 
\usepackage{subcaption}
\usepackage{hyperref}

\usepackage[accepted]{icml2024}

\usepackage{amsmath}
\usepackage{amssymb}
\usepackage{mathtools}
\usepackage{amsthm}
\usepackage{amsfonts}       
\usepackage{mathrsfs}
\usepackage[capitalize,noabbrev]{cleveref}

\theoremstyle{plain}

\theoremstyle{definition}

\theoremstyle{remark}

\newcommand{\R}{\mathbb{R}}

\newcommand{\M}{\mathscr{M}}

\usepackage[textsize=tiny]{todonotes}

\icmltitlerunning{Dynamic Basis Function Interpolation for Adaptive In Situ Data Integration in Ocean Modeling}

\begin{document}

\twocolumn[
\icmltitle{Dynamic Basis Function Interpolation for Adaptive In Situ Data Integration in Ocean Modeling}




\begin{icmlauthorlist}
\icmlauthor{Derek DeSantis}{LA}
\icmlauthor{Phillip J. Wolfram}{NG}
\icmlauthor{Ayan Biswas}{LA}
\icmlauthor{Earl Lawrence}{LA}
\end{icmlauthorlist}

\icmlaffiliation{LA}{Los Alamos National Laboratory, Los Alamos, USA}
\icmlaffiliation{NG}{Northrop Grumman, Boulder, CO, USA}


\icmlcorrespondingauthor{Derek DeSantis}{ddesantis@lanl.gov}

\icmlkeywords{Basis function interpolation, Data assimilation, dynamical systems learning,   ICML}

\vskip 0.3in
]



\printAffiliationsAndNotice{}  

\begin{abstract}
We propose a new method for combining in situ buoy measurements with Earth system models (ESMs) to improve the accuracy of temperature predictions in the ocean. The technique utilizes the dynamics \textit{and} modes identified in ESMs alongside buoy measurements to improve accuracy while preserving features such as seasonality. We use this technique, which we call Dynamic Basis Function Interpolation, to correct errors in localized temperature predictions made by the Model for Prediction Across Scales Ocean component (MPAS-O) with the Global Drifter Program's in situ ocean buoy dataset.
\end{abstract}

\section{Introduction}


In Earth System Sciences, improved accuracy of measurements and models is needed to better understand the climate. Powerful Earth system models (ESMs) provide estimates of evolution of state variables across grid cells, e.g., espacially at high resolution as in the Department of Energy’s  Energy Exascale Earth System Model (E3SM). E3SM's ocean component the Model for Prediction Across Scales Ocean (MPAS-O) evaluates Eulerian temperature and salinity across scales using multi-resolution hexagonal ocean cells  \cite{golaz2019doe,petersen2019evaluation,ringler2013multi}. Direct in situ observations, such as those collected by satellites, towers, or ocean buoys provide validation for model results. Although these observations are more accurate at specific locations, they are quite limited in coverage even for comprehensive experimental platforms. For instance,  the Global Drifter dataset program (GDP) overseen by NOAA \cite{elipot2022dataset,elipot2022hourly} has produced a highly resolved ocean buoy dataset. Simulation and observational approaches each have their advantages and disadvantages, and integrating both types of data can provide the greatest flexibility and predictive power \cite{reichle2008data}.


\textit{This work focuses on efficiently correcting Earth system model biases as measured through in situ observations through basis function interpolation (BFI). } Basis function models, which use linear combinations of simple functions to estimate a state, are a flexible and computationally efficient method for modeling non-stationary systems \cite{cressie2022basis}. In these models, a set of basis functions are chosen or discovered, and the associated coefficient weights for an unknown function in the space are learned.  In this paper, we propose a new \textit{dynamical} basis function interpolation (DBFI) for data integration based off the dynamic mode decomposition. This DBFI, which costs as little as a single SVD call, can correct ESM spatial temporal biases at arbitrary scales. In particular, the method can adjust for short term seasonality, or longer slower temporal climate modes of variability. This DBFI is deployed to analyze surface temperature from the MPAS-O model and the GDP to correct for MPAS-O Eulerian model biases.

As the name suggests, BFI and DBFI provides a global spatial interpolation of the in situ measurements using Earth system model measurements. This contrasts with previous work in this area predominantly focused on the (ensemble) Kalman filter (KF) and related techniques \cite{carton2008reanalysis}. Data assimilation using KF typically provides corrections to the model based on availability of in situ observations, with the state of a model in areas without direct observations updated based on the model's own dynamics and the influence from nearby corrected states. While known to reduce model errors quite well, KF assumes Gaussian error distributions, requires careful tuning of ensemble sizes, and can be quite expensive to run \footnote{Ensemble KF costs $\mathcal{O}(NLK)$ where $N$ is the number of in situ observations, $L$ is grid size, $K$ is size of ensemble \textit{per time step iteration}.} \cite{roth2017ensemble}. Further, the effectiveness of spreading updates from observed to unobserved areas depends heavily on the resolution of the model and the density of observations. Higher resolution models with sparse observations might exhibit less effective spread of updates compared to coarser models with the same observational density.  

This work focuses on the effectiveness of DBFI as a cost efficient alternative. DBFI provides a simultaneous global interpolation, cutting out the need to emulate dynamics with the model or assuming any particular noise model.  This can be done regardless of model fidelity, and can be done  significantly cheaper.

The paper is organized as follows. In Section Two, we provide a brief summary of the background information.  In Section Three, we present some results and in Section Four discuss how the dynamic method is able to correct for biases within the MPAS model while simultaneously adjusting for temporal variations, which static basis methods fail to capture.  The Appendix further details the methods and implementation.

\section{Methods}

\subsection{Basis Function Interpolation and Data Integration}
\label{Section: basis}

Throughout, we treat in situ observations as ground truth of the state, and ESM as an estimate. While it is known that GDP has biases, this work is not focused on correcting these biases, but rather using these estimates to calibrate our ESM.

Given a domain of interest $\M$ with $\{x_i\} \subset \M$ known (bouy) locations with corresponding (temperature) values $\{y_i\}$, interpolation involves finding a function $F:\M \rightarrow \mathbb{R}$ such that $F(x_i) \cong y_i$ for $i=1, \dots, N$. In \textit{basis function interpolation (BFI)}, $F$ is  represented  as a linear combination of simple \textit{basis functions} $\phi_j:\M \rightarrow \mathbb{R}$:
\[
F(x) = \sum_{j=1}^M a_j \phi_j(x).
\]
To perform BFI first the basis functions $\{\phi_j\}_{j=1}^M$ are selected or discovered.  Then a regression problem is solved:
\[
y_i \cong F(x_i) = \sum_{j=1}^M a_j \phi_j(x_i).
\]
This can be written in matrix form as
\begin{equation}
\label{regression}
\vec{y} \cong \Phi \vec{a},
\end{equation}
where $\Phi$ is the matrix with entries $\Phi_{i,j} = \phi_j(x_i)$, $\vec{y} = (y_1, \dots, y_N)^T$, and $\vec{a} = (a_1, \dots, a_M)^T$. The least-square solution to Equation \ref{regression} is obtained by taking the Moore-Penrose inverse of $\Phi$ on both sides:
\[
\vec{a} \cong \Phi^{\dagger} \vec{y}.
\]
The Moore-Penrose inverse can be computed using the singular value decomposition (SVD) of $\Phi$.  Often the matrix $\Phi$ is ill-conditioned, in which case regularization is needed. Optimization is performed over a range of regularization parameters to find the best fit on the training data. Once the coefficients $\vec{a}$ have been obtained, the model can be tested for accuracy on the test data.



\subsection{Static and Dynamic Basis Function Interpolation}
\label{Section: dynamic basis}

Basis functions $\{\phi_j\}_{j=1}^M $ are either handpicked or discovered from data.  To perform ESM and buoy data model assimilation through BFI, we need to discover the basis functions from the ESM.  There are several well established methods for extracting basis functions from model observations.  While the SVD is a popular method for extracting spatially coherent modes, the Dynamic Mode Decomposition (DMD) is an alternative method which emphasizes the discovery of explainable seasonality (See Appendix). 

The spatial mode from SVD or DMD can be used as the static basis functions as explained below. However both of the mode-based decompositions from SVD and DMD also include additional dynamical information. In this subsection, we present a method for incorporating this information into BFI to develop a true \textit{dynamical basis function interpolation (DBFI)}. Concretely, we seek basis functions $\phi_j(x,t)$ which depend on \textit{both} the space and the time the observation is taken:
\[
F(x,t) = \sum_{j=1}^M a_j \phi_j(x,t).
\]

Throughout, let $t_j$, $j=1,2,\dots T$ denote the discrete time observations for \textit{both} the in situ and model observations, and $\{w_l\}_{l=1}^L$ be the grid cell locations for our ESM over the domain $\M$. For each time $t \in [t_1:t_T]$, the ESM data ${z_l(t)}_{l=1}^L$ provides an estimate of the temperature distribution over the spatial domain $\M$ at time $t$, which can be arranged into an $L\times T$ matrix $Z$.

In SVD, the left singular vectors of $U = [U_1, \dots, U_M] \in \R^{L,M}$ represent spatial patterns, while the right singular vectors $V = [V_1, \dots, V_T] \in \R^{T,M}$ represent the intensity of each of the $M$ patterns across the $T$ time slices. Similarly, DMD produces its $M$ DMD modes $U = [U_1, \dots, U_M] \in \R^{L,M}$ and their associated dynamics (intensity across time) $V = [V_1, \dots, V_T] \in \R^{T,M}$ as described in the Appendix.

Let $\hat{x}$ denote the index of the ESM cell that location $x \in \M$ belongs to.  We define the \textbf{static (SVD or DMD) mode basis} as
\[
\phi_j(x):= U_{\hat{x}, j}
\]
for each $j=1, \dots, M$. In other words, the static mode basis is simply the spatial modes obtained through the modal decomposition (SVD or DMD).

Each in situ measurement $(x_i, y_i)$ is taken at a specific time $t \in [t_1:t_T]$. More generally, we can associate each point $x \in \M$ with a time $t \in [t_1:t_T]$ at which we want to provide interpolation.  We define the \textbf{dynamic (SVD or DMD) mode basis} as
\[
\phi_j(x,t):= V_{t,j} U_{\hat{x}, j}
\]
for each $j=1, \dots, M$. In other words, the dynamic mode basis is obtained from the static mode basis by weighting the static mode at an observation by its intensity at the time of the observation. This dynamic weighting accounts for the evolving intensity of modes over time, adjusting for seasonality and any other cycles (e.g. slower/faster modes of variability).

We note that since the dynamic information ${V_{t,j}}$ has already been computed, the added complexity compared to the static method is $\mathcal{O}(1)$ per observation. Thus the total cost of DBFI is approximately the same as BFI, roughly costing a \textit{single} regression and mode extraction ($\mathcal{O}(NMR)$ rank $R$ regression/top $R$ modes). In many regimes, this is \textit{significantly} cheaper than KF methods. 

\subsection{Model Summary}
We investigate the effect of adding dynamics to the BFI. As a baseline, we will compare the performance of our basis models to the ESM.  We will also compare the differences BFI and our DBFI. Table \ref{tab:models} provides the summary of the models in this short note.

\begin{table}[h!]
\resizebox{.5\textwidth}{!}{%
\begin{tabular}{ll}
\textbf{Model} & \textbf{Note}                                          \\
MPAS           & ESM, baseline comparison                               \\
SVD            & SVD discovered $\phi_j$, static modes                  \\
DMD            & DMD discovered $\phi_j$, static modes                  \\
SVD\_dyn       & SVD discovered modes with added dynamics               \\
DMD\_dyn       & DMD discovered modes with added dynamics              
\end{tabular}%
}
\caption{Overview of models.  MPAS provides a baseline to measure improvements with data integration.  SVD and DMD give a control to compare by adding dynamics. }
\label{tab:models}
\end{table}




\section{Results}



We investigate sea surface temperature DBFI using MPAS-O as our ESM and the GDP for ocean buoy data. Specifically, the physical domain of interest $\M$ will denote the Atlantic Ocean between latitudes $15^\circ-55^\circ$~N and longitudes $260^\circ-20^\circ$~E. We will be using the V1 historical run of MPAS-O, which has ocean surface resolution $6-18 \mbox{km}$, while the  GDP sea surface buoy measurements are from 1990 onward \cite{ringler2013multi,petersen2019evaluation,doecode_10475,elipot2022dataset}. These two datasets have different temporal resolutions, so some time averaging within the GDP is required to match the MPAS-O. See the appendix for more details.

\begin{figure}[h!]
    \centering
    \includegraphics[width=\linewidth]{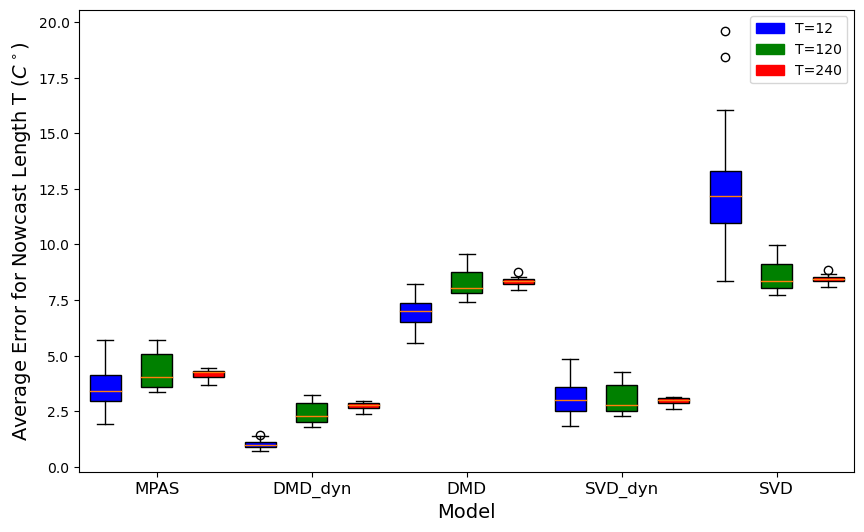}
    \caption{ Bias error distribution of models in Table \ref{tab:models}. Each model is fit on $80\%$ of the data and tested on the remaining $20\%$ to produce an error measurement. The distribution of errors for each model is created by choosing different starting times $t_1$ and nowcast time horizions $T$ (in months).  }
    \label{fig: error_wiskers}
\end{figure}

 Several time horizons $T$ were investigated for now-cast predictions.   Models were fit using $80\%$ of available in situ data for the time horizon, and tested on the remaining $20\%$ (chosen uniformly). Figure \ref{fig: error_wiskers}  displays the distribution test errors for initializing the  models at different starting times. We note that the DMD\_dyn model outperforms all other models by sizeable margin, and the non dynamic models fail to improve upon the MPAS baseline.

\begin{figure}[ht]
\centering
\begin{subfigure}{0.23\textwidth}
    \includegraphics[width=\linewidth]{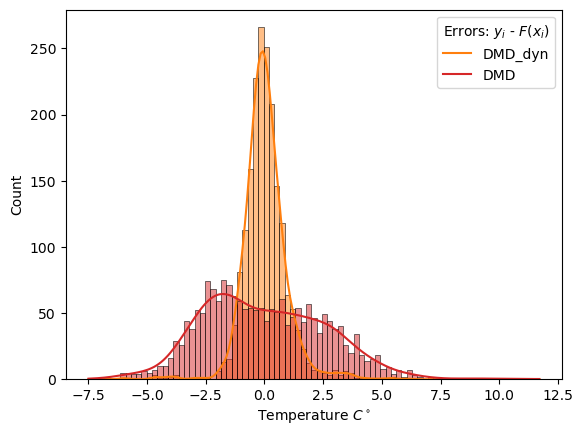}
    \caption{Bias errors of DMD\_dyn basis versus adding dynamics}
\end{subfigure}
\hfill
\begin{subfigure}{0.23\textwidth}
    \includegraphics[width=\linewidth]{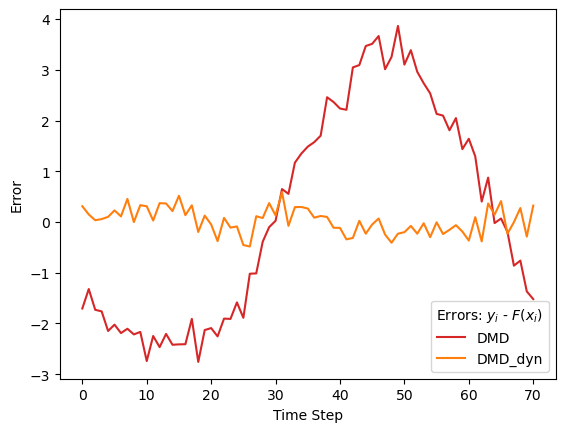}
    \caption{Bias error over time - DMD basis versus adding dynamics}
\end{subfigure}


\begin{subfigure}{0.23\textwidth}
    \includegraphics[width=\linewidth]{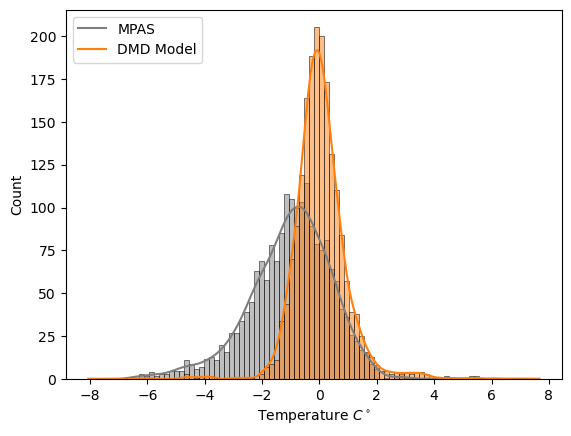}
    \caption{Bias errors of DMD\_dyn basis ESM}
\end{subfigure}
\hfill
\begin{subfigure}{0.23\textwidth}
    \includegraphics[width=\linewidth]{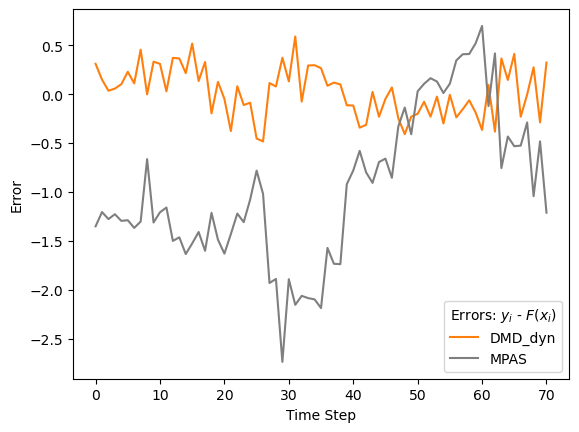}
    \caption{Bias error over time - DMD\_dyn basis versus ESM}
\end{subfigure}

\caption{Bias errors of models for time period Jan. 1st 2008-Jan 1st 2009.  Plots (a) and (c) display error distributions for different models.  Plots (b) and (d) compute the average error on each of the five day time slices to explore inability to capture seasonality/baises.}
\label{fig: error_dist}
\end{figure}

Next, we will perform a deeper analysis of the model performance over a single year to illuminate how the DMD\_dyn models improved performance manifests. For demonstration purposes, we have selected the date range of Jan. 1st, 2008 to Jan 1st, 2009 to display here \footnote{While we focus on Jan. 1st, 2008 to Jan 1st, 2009, the results presented here are representative of the general year-long phenomenon. Similar analysis can be applied to any horizon $T$.}. Figure \ref{fig: error_dist}(a) shows how the test error is improved by adding the dynamic component to the DMD mode decomposition. The DMD basis decomposition has a somewhat bimodal error, indicating that the model has over- or undercompensated for SST in some regions. Figure \ref{fig: error_dist}(b) illustrates that this is due to seasonality - the basic DMD model has discovered a mean state to represent the temperature, missing the extremes of both summer and winter. Weighting the modes by their dynamics removes these biases, producing a normally distributed error with a much lower standard deviation. Figure \ref{fig: error_dist}(c) shows that MPAS has a hot temperature bias, pushing the distribution's mean into the negatives. This is further seen in Figure \ref{fig: error_dist}(d), where MPAS appears to have a negative temperature bias across the whole year, which has been corrected in DMD\_dyn.

\begin{figure}[ht]
\centering
\begin{subfigure}{0.5\textwidth}
    \includegraphics[width=\linewidth, height=2.5cm]{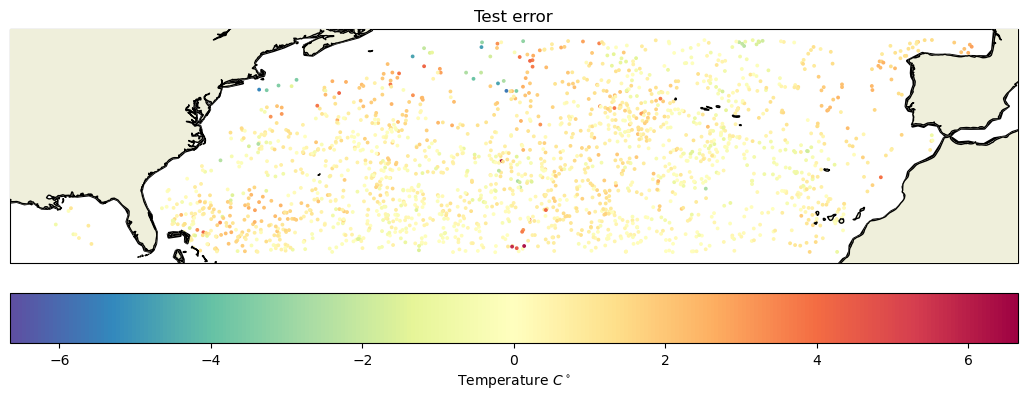}
    \caption{MPAS errors on buoys}
    \label{fig:image1}
\end{subfigure}
\hfill 
\begin{subfigure}{0.5\textwidth}
    \includegraphics[width=\linewidth, height=2.5cm]{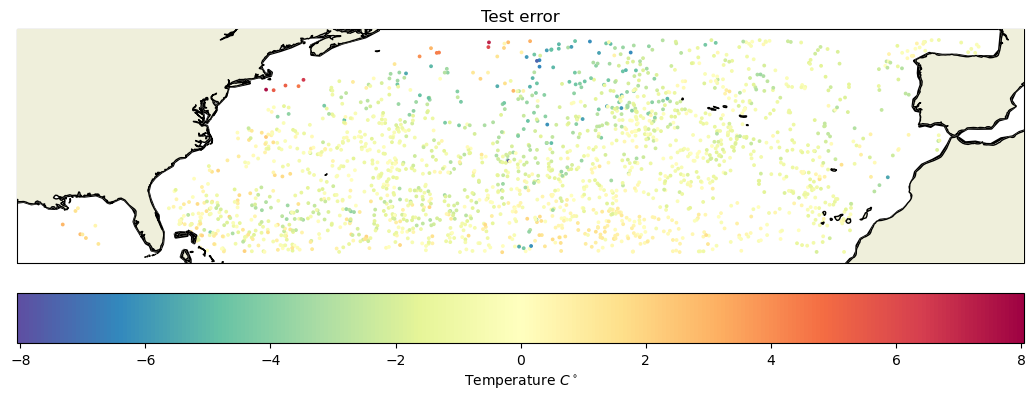}
    \caption{DMD\_dyn errors on buoys}
    \label{fig:image2}
\end{subfigure}
\hfill 
\begin{subfigure}{0.5\textwidth}
    \includegraphics[width=\linewidth, height=2.5cm]{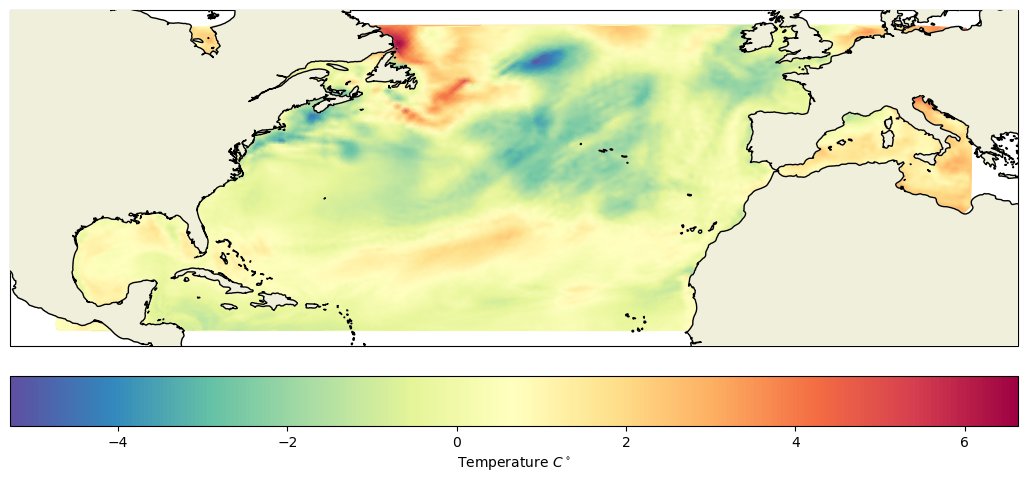}
    \caption{MPAS bias errors on buoys: $y_i - F(x_i)$}
    \label{fig:image3}
\end{subfigure}

\caption{Spatial comparison of DMD\_dyn and MPAS.}
\label{fig:spatial plotts}
\end{figure}

Lastly, we  analyze the spatial distribution of DMD\_dyn interpolation. Figure \ref{fig:spatial plotts}(a) and (b) plot the test errors of DMD\_dyn and MPAS on the actual buoys.   The buoy errors show that MPAS has a clear latitudinal temperature bias, whereas the fully dynamic DMD has relatively uniform errors. Figure \ref{fig:spatial plotts}(c) shows the difference between the DMD\_dyn and MPAS averaged over the full year. The DMD\_dyn has created a physical model similar to MPAS-O, providing temperature corrections to MPAS-O in the appropriate regions.


\section{Discussion}


ESMs have intrinsic model biases.  As noted, MPAS-O has a latitudinal temperature bias in its V1 historical simulation \cite{golaz2022doe}.  The simple BFI do not improve the accuracy of the ESM.  However it is important to note that DMD provides better estimates of spatio-temporal evolution than SVD. This could be due in part to the fact that DMD modes have dynamics described by the growth and decay of eigenvalues. As such, the spatial modes are typically suggestive with particular months, seasons, cycles, etc. Consequently, DMD can intrinsically adapt to seasonal changes better. While SVD does provide dynamic information, DMD is explicitly built to extract different modes of variability.

\textbf{The key discovery from the previous section is that adding available dynamics, specifically to the DMD method of interpolation, improves predictive skill (Figure \ref{fig: error_wiskers}) while correcting for seasonality (Figure \ref{fig: error_dist})}.  This is because the  DMD\_dyn model captures different time scales and reweights them into the model.  DBFI improve the biases of MPAS-O by assimilating the sparse in situ bouy measurements into the model (Figure \ref{fig:spatial plotts}).  Notably, the DMD\_dyn model corrects the latitudinal temperature bias.  The discovered DMD modes capture the dynamics of MPAS-O, while having the flexibility to be reweighted to match the real observational data.



\section{Conclusion}

In this paper, we introduced a dynamic basis function interpolation in order to 1) Leverage in situ observations to correct ESM biases,  2)  Dynamically weight the influence of basis functions over window of investigation in way that is 3) Fast and 4) Flexible to any temporal horizon. We found that adding dynamic information, specifically to the DMD basis functions, improves predictive accuracy and corrects for biases on longer time scales with seasonal dynamics. DMD\_dyn  was consistently the best performer among all methods. We demonstrated that DMD\_dyn is not overfitting and is making reasonable spatial predictions while also correcting for biases in MPAS-O. This type of dynamic decomposition for interpolating ocean data that can adapt to both high and low frequency information show great potential.  The dynamic data assimilation technique discussed here allow for more  data-rich analysis that could be useful for understanding evolving dynamics.  Given their natural ability to scale and adapt to  online regimes, further development in that direction appears most promising.

%
%

\bibliography{example_paper}
\bibliographystyle{icml2024}

\newpage
\appendix
\onecolumn
\section{Appendix}

\subsection{Time Filtering GDD Data to MPAS-O}

The MPAS-O and GDP datasets are on different temporal resolutions (Five day averages versus six hour). We therefore coarse-grained the GDP data in time to fit the temporal resolution of MPAS. Given two sequential MPAS-O time snapshots $t$ and $t+1$, all the GDP buoy data is collected within each of those five days.  For each buoy, the average spatial location and temperature is then recorded and used as $\{x_i(t)\}$ and $\{y_i(t)\}$.






\subsection{DMD Short Summary}

In the Dynamic Mode Decomposition (DMD), the system is assumed to be evolved in an approximately linear fashion:
\begin{equation}
    \label{DMD single}
\vec{z}_{t+1} \cong A \vec{z}_t.
\end{equation}
This is effectively the solution of a forward-in-time spatially discrete solver, where $A$ can contain physical operators such as advection, diffusion, etc.  The value in this form is that growth, decay, and oscillatory behavior is all immediately represented, albeit pseudo-linearly from one time step to the next.  Given this interpretation, one can think of $A$ as effectively a learned, empirical physically-meaningful discretization.

Let $Z_1$ and $Z_2$ be the matrices given by $Z_1 = [\vec{z}_1 \vec{z}_2 \dots \vec{z}_{T-1}]$ and $Z_2 = [\vec{z}_2 \vec{z}_3 \dots \vec{z}_{T}]$.  The goal is to find a good approximation $\tilde{A}$ of the matrix $A$ that represents this system. This  can be cast as the following matrix problem:
\begin{equation}
    \label{DMD}
    Z_2 \cong A Z_1
\end{equation}
The least square solution to Equation \ref{DMD} is found by taking the Moore-Penrose inverse:
\[
A \cong Z_2 Z_1^\dagger.
\]
Since $A$ is a square matrix, one can consider the eigen-decomposition of $A$:
\begin{equation}
    \label{DMD modes}
    A U = U \Lambda
\end{equation}
The eigenvectors and eigenvalues $(u_j,\lambda_j)$ of the $A$ are the \textbf{DMD modes and eigenvalues} respectively.

For systems with a large number of data points $L$, this direct method isn't tractable since $A$ is a $L \times L$ square matrix.  Therefore different methods are designed to 1) solve Equation \ref{DMD} while 2) providing a reduced order model at the same time. Among the most simple and popular methods for doing this is the SVD-based method:

\begin{enumerate}
    \item Compute the rank $M$ SVD of $Z_1 = W \Sigma V^T$.
    \item Consider $\tilde{A}:=W^T A W$.  Then $\tilde{A}$ is $M\times M$, and since it is unitarily equivalent to $A$, has the same eigenvalues with eigenvectors $\xi$ of $\tilde{A}$ related to $A$ via $W \xi$.
    \item Note that since $Z_2 \cong A Z_1 = A W \Sigma V^T$, after multiplying by $W^T$ and rearranging we have
    \[
    \tilde{A} = W^T A W = W^T Z_2 V \Sigma^{-1}.
    \]
    In other words, $\tilde{A}$ can be computed directly from the data and SVD of $Z_1$. 
    \item Compute eigenvectors and eigenvalues $\{\tilde{u}_j,\lambda_j\}$ of the much smaller $M \times M$ matrix $\tilde{A}$.
    \item The $j'$th DMD mode is computed as $u_j:=W \tilde{u}_j$.
\end{enumerate}

Other methods for approximating the DMD modes based off SVD exist, such as exact DMD.  See \cite{tu2013dynamic}.

\begin{figure}[ht]
    \centering
    \begin{subfigure}[b]{0.45\linewidth}
        \includegraphics[width=\linewidth]{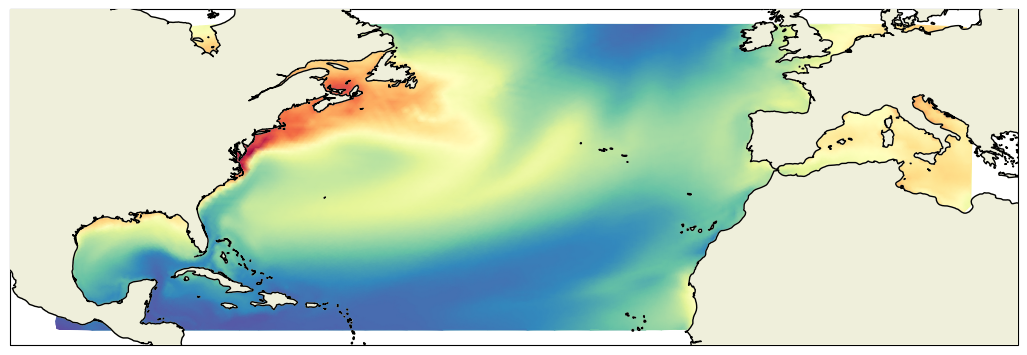}
    \end{subfigure}
    \hfill
    \begin{subfigure}[b]{0.45\linewidth}
        \includegraphics[width=\linewidth]{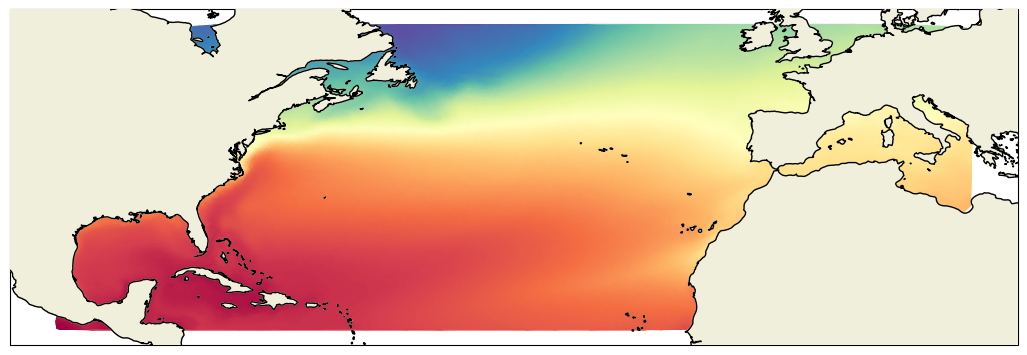}
    \end{subfigure}
    
    \begin{subfigure}[b]{0.45\linewidth}
        \includegraphics[width=\linewidth]{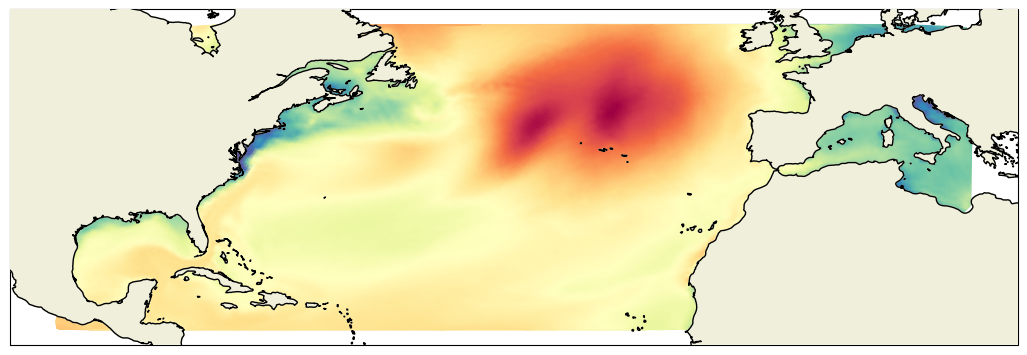}
    \end{subfigure}
    \hfill
    \begin{subfigure}[b]{0.45\linewidth}
        \includegraphics[width=\linewidth]{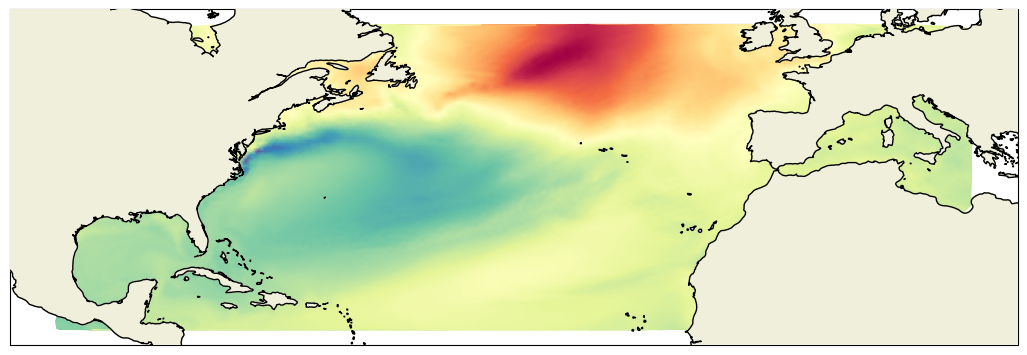}
    \end{subfigure}
    \caption{DMD Mode plots.}
    \label{fig:dmd_modes}
\end{figure}

  The DMD modes and eigenvalues can be used to derive dynamical information analogous with the right singular vectors of SVD. By iteratively applying Equation \ref{DMD single}, we find that 
\begin{equation}
    \label{DMD Dynamics}
    \vec{z}_{t+1} \cong A^{t} \vec{z}_1
\end{equation}
Write $\vec{z}_1$ in the basis provided by the modes $U$:
\begin{equation}
    \label{DMD init}
    \vec{z}_1 = U\vec{b}.
\end{equation}
Then combining Equations \ref{DMD modes}, \ref{DMD Dynamics} and \ref{DMD init} we see that
\[
\vec{z}_{t+1} \cong A^{t} \vec{z}_1 =  A^{t}  U \vec{b} =  U \Lambda^{t} \vec{b}.
\]
Hence, we have represented the state of the system in terms of a DMD mode expansion with temporal evolution on the DMD eigenvalues. The time series $V_{t,j}:=\lambda_j^t b_j$ for $t=1,\dots T$ and $j=1,\dots M$ are referred to as the \textbf{DMD dynamics} for the $j$'th mode $u_j$. Figure \ref{fig:dmd_modes} plots example spatial modes $u_j$ of a DMD decomposition for the MPAS-O dataset over the Atlantic Ocean between latitudes $15^\circ-55^\circ$~N and longitudes $260^\circ-20^\circ$~E.

\end{document}